\documentclass[journal]{IEEEtran}
\usepackage{multirow}
\usepackage[justification=centering, font=footnotesize]{caption}
\usepackage{cite}

\ifCLASSINFOpdf
  \usepackage[pdftex]{graphicx}
  \graphicspath{{../pdf/}{../jpeg/}}
  \DeclareGraphicsExtensions{.pdf,.jpeg,.png}
\else
  \usepackage[dvips]{graphicx}
  \graphicspath{{../eps/}}
  \DeclareGraphicsExtensions{.eps}
\fi
\usepackage[hidelinks]{hyperref}

\usepackage{amsmath}

\usepackage{booktabs}
\usepackage{amssymb}

\begin{document}
%
\title{End-to-End Performance of Video Streaming With MPEG-DASH Over Satellite 5G IAB Networks}
\author{Muhammad Adeel Zahid, Ekram~Hossain,~\IEEEmembership{Fellow,~IEEE}, and Peng~Hu,~\IEEEmembership{Senior Member,~IEEE}\thanks{The authors are with the
Department of Electrical and Computer Engineering,
University of Manitoba, Winnipeg, Canada 
(Emails: zahidma@myumanitoba.ca, Ekram.Hossain@umanitoba.ca, Peng.Hu@umanitoba.ca)}
}
\maketitle

\begin{abstract}
We present an end-to-end performance evaluation of MPEG-DASH video streaming over a Low-Earth Orbit (LEO) satellite-based 5G Integrated Access and Backhaul (IAB) network. Our objective is to investigate how modern transport protocols and congestion control algorithms affect adaptive video delivery in an integrated satellite-terrestrial network (ISTN), where latency, throughput variation, and playback continuity jointly shape the user Quality-of-Experience (QoE). We implement a simulation framework in ns-3 by adapting open-source modules for the 5G radio access network, LEOS backhaul, transport layer protocols, and MPEG-DASH application behavior. Within this framework, TCP and QUIC are evaluated with multiple congestion control algorithms, including CUBIC, NewReno, and BBR. Performance is assessed using application-level and transport-level metrics, including playback duration, interruption duration, stall count, playback bitrate, throughput, latency, and fairness. The results show that no single configuration is uniformly optimal across all metrics. However, clear tradeoffs are observed among throughput, latency, playback continuity, and fairness. In particular, QUIC-BBR provides the most balanced overall behavior from a streaming QoE perspective, combining adequate playback duration with fewer interruptions and substantially lower latency than other alternatives. These findings highlight the importance of jointly considering transport design and congestion control when evaluating adaptive video streaming over ISTNs.
\end{abstract}


%
\IEEEpeerreviewmaketitle

\section*{Introduction}
Video streaming is one of the most demanding and dominant forms of Internet traffic, and its performance depends strongly on the interaction between the application, transport, and underlying network. For adaptive streaming systems such as MPEG-DASH, quality-of-experience (QoE) is shaped not only by available throughput, but also by latency, packet loss, playback interruptions, and bitrate stability. Therefore, evaluating video streaming performance requires an end-to-end perspective that captures both network behavior and application-level outcomes.

Currently, the integration of terrestrial 5G systems with low-Earth orbit (LEO) satellite networks is creating a new environment for broadband service delivery. Compared with conventional terrestrial backhaul, LEOS networks introduce propagation delay, dynamic link conditions, and time-varying transport characteristics. These properties make such systems an important setting for understanding how modern transport protocols behave under realistic but challenging operating conditions.

Although prior work~\cite{Kim2019,Ge2019,Wang2018,Fang2024,Alshagri2023,Yang2018} has studied satellite-terrestrial integration, transport performance over satellite networks, and adaptive streaming mechanisms separately, fewer studies have examined their interaction in a unified end-to-end framework that jointly captures transport-level behavior and QoE. In~\cite{Ge2019}, multiaccess edge computing (MEC)-assisted satellite backhaul and prefetching are considered for QoE support in 5G streaming, without addressing the characteristics of the IAB architecture. Other representative works examine transport behavior in ISTN~\cite{Alshagri2023,Yang2018}, handover-aware live streaming over LEO satellite constellations~\cite{Fang2024}, and broader architectural or delivery-support mechanisms for satellite-terrestrial integration~\cite{Kim2019,Wang2018}.  

5G IAB was first standardized in 3GPP Release 16, which allows the 5G New Radio (NR) spectrum to serve both UE access links and wireless backhaul among gNBs. However, current 3GPP standardization efforts neither address IAB architectures employing LEO satellite backhaul nor consider performance evaluations for key video streaming applications.

To address the gap in the existing literature, this paper presents an end-to-end performance analysis of MPEG-DASH video streaming over a LEO-backhauled 5G integrated access and backhaul (IAB) network, enabling joint evaluation of transport protocol choice, congestion control behavior, and application-level QoE. This is based on a simulation framework implemented in ns-3 by customizing and integrating open-source modules for the radio access, satellite backhaul, transport, and application layers. Within this framework, we compare TCP and QUIC under multiple congestion control algorithms and evaluate their impact on playback continuity, bitrate delivery, throughput, latency, and fairness. The results show that the choice of transport protocol and congestion control algorithm leads to clear tradeoffs across these metrics. Rather than identifying one configuration as uniformly superior in every respect, the study highlights how different schemes balance responsiveness, continuity, throughput, and fairness in distinct ways. This provides a more complete understanding of protocol behavior for video streaming over an ISTN. Table~\ref{tab:related_work_comparison} summarizes the scope of these representative studies and highlights the distinctions of the present work.

\begin{table*}[t]
\centering
\caption{Comparison with related work}
\label{tab:related_work_comparison}
\renewcommand{\arraystretch}{1.15}
\setlength{\tabcolsep}{6pt}
\begin{tabular}{lccccc}
\toprule
\textbf{Work} & \textbf{ISTN} & \textbf{Transport layer Analysis} & \textbf{QoE} & \textbf{IAB Architecture} & \textbf{End-to-End Analysis} \\
\midrule
Kim \textit{et al.}~\cite{Kim2019} & $\checkmark$ &  &  &  &  \\
Ge \textit{et al.}~\cite{Ge2019} & $\checkmark$ &  & $\checkmark$ &  &  \\
Wang \textit{et al.}~\cite{Wang2018} & $\checkmark$ &  & $\checkmark$ &  &  \\
Alshagri \textit{et al.}~\cite{Alshagri2023} &  & $\checkmark$ & $\checkmark$ &  &  \\
Yang \textit{et al.}~\cite{Yang2018} & $\checkmark$ & $\checkmark$ & $\checkmark$ &  &  \\
Fang \textit{et al.}~\cite{Fang2024} &  &  & $\checkmark$ &  &  \\
\textbf{This work} & $\checkmark$ & $\checkmark$ & $\checkmark$ & $\checkmark$ & $\checkmark$ \\
\bottomrule
\end{tabular}
\end{table*}

\section*{Background and Related Work}

The literature relevant to our work spans three critical and converging areas: the architectural design of ISTNs, the performance of transport layer protocols over high-latency links, and the assurance of QoE for adaptive video streaming.

The evolution of 5G architecture has been moving toward integration of non-terrestrial networks (NTNs) and terrestrial infrastructure to provide enhanced mobile broadband in remote and under-served areas. The authors in \cite{Kim2019} focus on Multi-Connectivity (MC) to seamlessly integrate cellular and satellite links. MC is a core 5G feature, standardized primarily in 3GPP Release 15 and enhanced in later releases, that allows a device to simultaneously communicate over two different wireless technologies. In this integrated system, MC is crucial for combining the high-capacity, local access of 5G New Radio (NR) with the wide-area coverage of the non-terrestrial satellite links. The terrestrial component of this link assumes connectivity via standard 5G base stations.

The authors in \cite{Ge2019} propose deploying Multi-access Edge Computing (MEC) servers at the 5G edge to offload content traffic via satellite backhaul. This strategy effectively breaks the end-to-end path into two distinct segments: the satellite backhaul and the 5G Radio Access Network. By localizing content delivery at the MEC, the system uses parallel TCP connections to prefetch video segments before they are requested by the user. This approach masks the high latency of satellite links and ensures a stalling-free experience with minimal startup delays \cite{Ge2019, Wang2018}. 
The unique network environment of satellite links characterized by high latency, potential high packet loss, and dynamic connections, is the primary challenge. While Geostationary Earth Orbit (GEO) satellites impose RTTs exceeding 500~ms, LEO systems introduce frequent satellite handover events that cause abrupt network disruptions \cite{Fang2024}.


At the transport layer, performance of QUIC and TCP have been evaluated. QUIC, which was initially developed by Google and later standardized by the IETF as the transport basis of HTTP/3, is a UDP-based transport protocol designed for modern web and media delivery. It has been widely adopted for latency-sensitive services, such as Google Search and YouTube, because it reduces the delay in connection establishment, improves loss recovery, and eliminates transport-layer head-of-line blocking through native stream multiplexing. These properties make QUIC particularly attractive in high-latency environments such as satellite networks \cite{Alshagri2023, Yang2018}. Prior studies have shown that HTTP/3 over QUIC can outperform HTTP/2 over TCP in ISTN, especially in terms of page load time and transport efficiency \cite{Yang2018}. In addition to the transport protocol itself, the choice of congestion control algorithm is critical. In this context, the BBR algorithm is a model-based congestion control algorithm that estimates the available bottleneck bandwidth and minimum round-trip delay in order to regulate sending rate and the amount of in-flight data, rather than relying primarily on packet loss as a congestion signal \cite{IETFBBR}. This design generally allows BBR to achieve faster convergence and better link utilization than loss-based schemes such as CUBIC on large bandwidth-delay product satellite links \cite{Alshagri2023}. Even when TCP is enhanced with Explicit Congestion Notification (ECN) to signal impending congestion before packet loss occurs, QUIC has still been shown to provide superior performance in space-network scenarios due to its lower handshake latency, improved loss recovery, and more flexible transport architecture \cite{Yang2018, Alshagri2023}.

Despite its architectural advantages, QUIC is not without limitations. It is observed in \cite{Alshagri2023} that the added complexity of QUIC and HTTP/3 introduces protocol overhead that can become a dominant factor affecting performance, and consequently the QoE. This effect is especially noticeable when QUIC operates with slower-reacting congestion control algorithms such as CUBIC, in which case the additional overhead may reduce or even outweigh the gains offered by faster connection setup and improved loss recovery. Fairness is another important concern when QUIC operates with the BBR congestion control algorithm. Because BBR is a model-based congestion control algorithm, it regulates transmission using estimates of bottleneck bandwidth and minimum RTT rather than relying primarily on packet loss. This design can create unfairness in two cases. The first occurs when BBR competes with traditional loss-based algorithms such as CUBIC and NewReno. Since loss-based flows expand their congestion window until losses occur, they tend to build queues, whereas BBR attempts to pace traffic according to its bandwidth and delay model. The mismatch between these control principles can lead to unequal bandwidth sharing, with the outcome depending on factors such as buffer size and the aggressiveness of the competing flow. The second case arises when multiple BBR flows share a bottleneck but experience different RTTs. Under these conditions, the longer-RTT flow may capture more bandwidth because its larger bandwidth-delay product allows more in-flight data, while its slower control loop helps preserve an aggressive sending state once bandwidth is obtained. As discussed in \cite{Yue2025, Tang2024}, both mixed-algorithm competition and heterogeneous-RTT BBR coexistence can produce persistent fairness imbalances, even though BBR often maintains high throughput and low latency.

For the application layer, MPEG-DASH enables adaptive bitrate streaming, crucial for maintaining QoE. However, LEO satellite handovers lead to abrupt disruptions, causing significant video pauses that can last from a few seconds to more than twenty seconds \cite{Fang2024}. Effective solutions require the ABR (Adaptive Bitrate) logic to choose the best video quality (bitrate) at any moment, to be handover-aware to anticipate and mitigate the predictable timing of LEO handovers \cite{Zhao2024, Fang2024}. Techniques such as transient segment holding and prefetching, executed at the MEC server using multiple parallel TCP connections, are validated strategies to compensate for high latency \cite{Ge2019}. This enables stall-free streaming while minimizing live latency \cite{Ge2019, Zhao2024}.

\begin{figure*}[!t]
    \centering
    \includegraphics[width=\textwidth]{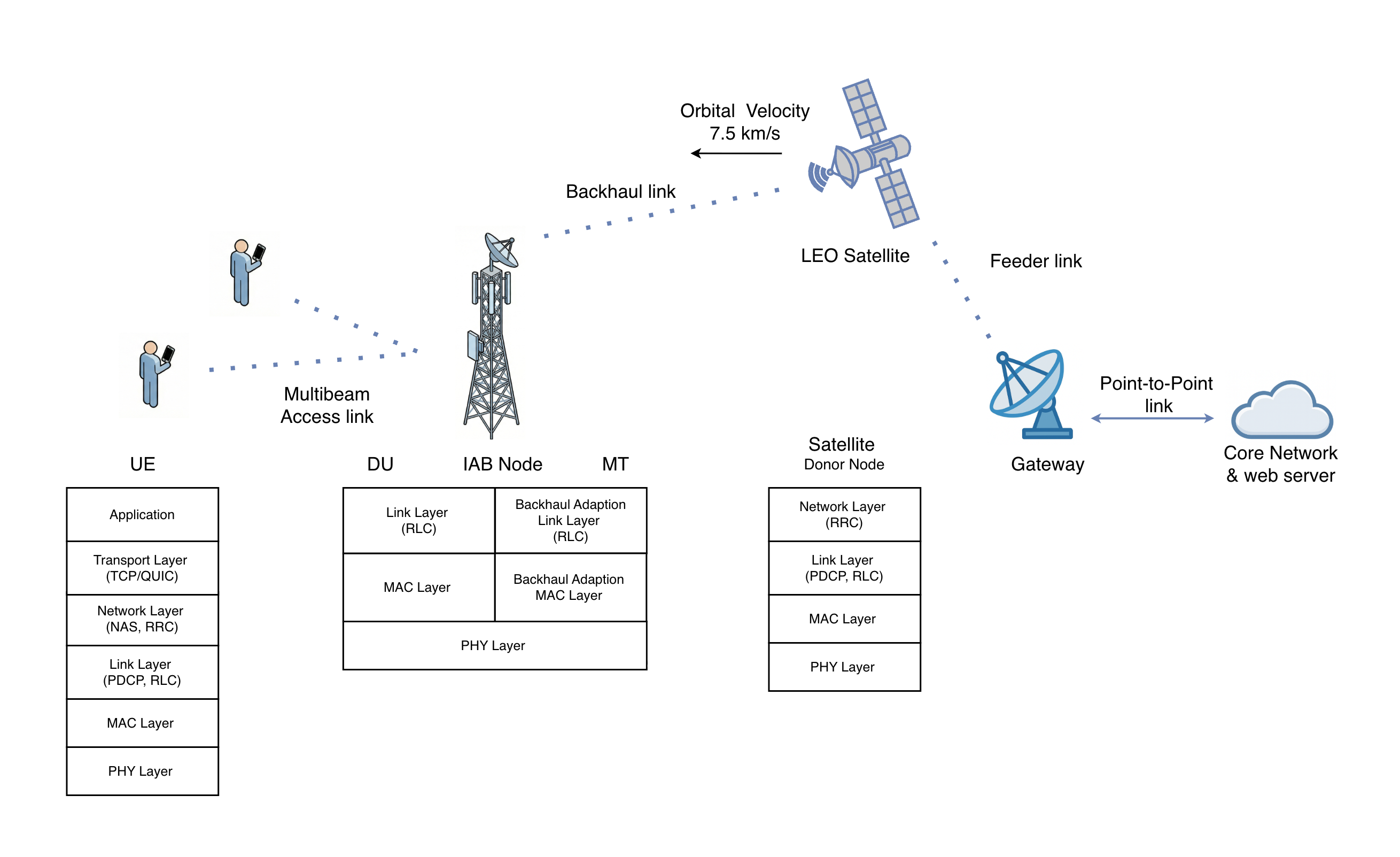}
    \caption{System architecture}
    \label{fig:system_architecture}
\end{figure*}

\section*{Network Architecture and System Overview}

The network architecture considered in this paper integrates 5G access with a non-terrestrial backhaul component for video streaming over MPEG-DASH. Fig. \ref{fig:system_architecture} illustrates the end-to-end topology, where the terrestrial IAB-node functions as a gateway for the terrestrial users. The number and placement of IAB nodes significantly impact the feasibility of the considered topology. Under realistic LEO satellite assumptions considered in this work, the system without an IAB relay remains in outage, as the direct link budget is not sufficient to provide reliable service to the user. The inclusion of an IAB node addresses this limitation by enabling a feasible terrestrial access link while utilizing the satellite connection for backhaul. Increasing the number of IAB nodes beyond one, however, leads to a degradation in link quality rather than an improvement. This is due to the additional interference introduced by the extra IAB nodes on the reception of the satellite signal. Moreover, because the satellite operates as a multibeam system, inter-beam interference further contributes to signal degradation. The resulting increase in losses can cause some users to lose connectivity entirely. Therefore, the system considered in this paper employs a single IAB node, placed centrally among the users.

The system design is modeled based on the five-layer Internet protocol stack and follows the 3GPP guidelines for cellular and radio communication and RFC standards for transport and application layer protocol. The simulation platform, i.e.,  \textbf{ns-3}, includes a number of open-source modules for various components out of which some are part of the open-source code directory. These modules are used in this work with certain modifications. Combining these community models, our system model evaluates the performance metrics of video streaming over two major transport protocols (TCP and QUIC) with a number of congestion control algorithms (BBR, NewReno, CUBIC). {\em The source code for this work is available at} \url{https://github.com/muhammadadeelzahid/ns3-ntn-iab}.

\subsection*{Application Layer}
Video streaming in the proposed system is implemented using MPEG-DASH, specified by ISO/IEC 23009. Adaptive bitrate streaming is a fundamental technique used by modern video protocols like DASH to cope with variable network conditions. It works by detecting a user's available bandwidth and device capabilities in real time and adjusting the quality of the video stream accordingly. In ns-3, we employ the MPEG-DASH adaptation module proposed in \cite{vergados2016fdash}, referred to as FDASH. The FDASH algorithm extends the standard MPEG-DASH rate adaptation process by introducing a fuzzy logic controller that uses the client’s instantaneous buffering time and its rate of change as inputs to determine the most appropriate bitrate. The video stream is divided into a number of segments, each consisting of multiple frames, encoded at different available bitrates. The controller first maps these two inputs into linguistic variables (short, close, long for buffer level, and falling, steady, rising for its change), which are then processed through a fuzzy rule base to infer whether the client should increase, decrease, or maintain the current video bitrate. The output of this fuzzy inference system is a scaling factor that adjusts the bitrate estimate derived from recent segment throughput measurements. FDASH is specifically designed to minimize resolution oscillations, avoid buffer underflow, and maintain stable playback quality under variable network conditions. 

\subsection*{Transport Layer}

At the transport layer, we focus on the interaction between the adaptive video application and the congestion control mechanisms that govern the LEO-backhauled 5G IAB path. To this end, we implement and compare two alternative transports: a traditional HTTP-over-TCP stack and a QUIC-based stack that represents HTTP/3 style operation. 

QUIC is a user-space transport protocol that runs over UDP and integrates cryptographic handshake, stream multiplexing, and loss recovery into a single layer. Standardized by the IETF in 2021 as RFC 9000, QUIC has since become the underlying transport for HTTP/3. Unlike TCP, which exposes a single ordered byte stream and therefore suffers from HOL blocking at the connection level, QUIC supports multiple independent streams. Loss on one stream does not block data delivery on others, which is particularly important in our setting where MPEG-DASH issues a sequence of HTTP requests and responses that can be mapped to different streams. QUIC also uses its own packet number spaces and rich acknowledgment information, which enables more precise loss detection and supports advanced congestion control and pacing strategies, especially over high-delay and variable-quality satellite links. In both cases, each MPEG-DASH client maintains a single long-lived connection to the video server and all segment requests are serialized over this connection, so that differences in performance are attributable to the transport rather than to connection management artifacts.

For the TCP-based experiments, we use ns-3's native TCP implementation configured with modern loss recovery and congestion control options. For QUIC, we extend the ns-3 module originally developed in \cite{DeBiasio2019}, which has undergone a series of upgrades including the implementation of the BBR congestion control algorithm.

\subsection*{Cellular Stack}

5G NR is the global standard for the 5G air interface, designed to support diverse use cases ranging from enhanced Mobile Broadband to Ultra-Reliable Low-Latency Communications. Key architectural advancements include support for millimeter-wave (mmWave) spectrum, which offers vast bandwidth resources, and flexible numerologies that adapt subcarrier spacing to different channel conditions. To overcome the high path-loss associated with mmWave frequencies, 5G NR heavily relies on massive MIMO and beamforming technologies to maintain robust link budgets.

The 5G stack, particularly with IAB, is simulated using the ns-3 mmWave module extended by \cite{Polese2018}. This module is further enhanced to incorporate the NTN channel and propagation models based on the work of \cite{Sandri2023}, which implements the 3GPP TR 38.811 channel model for satellite communication. Additionally, the system implements hybrid beamforming with multilayer functionality as described in \cite{Polese2020HBF}, enabling spatial multiplexing to communicate with multiple users simultaneously within the same time slot. This increases spectral efficiency and allows the network to support higher aggregate throughput.

\section*{Performance Evaluation}

\begin{table}[t]
\centering
\caption{Transport protocols and congestion control algorithms evaluated, and simulation parameters}
\label{tab:simulation_parameters}

\renewcommand{\arraystretch}{1.18}
\setlength{\tabcolsep}{6pt}

\begin{tabular}{|p{0.54\columnwidth}|p{0.38\columnwidth}|}
\hline
\textbf{Parameter} & \textbf{Value} \\ \hline

\multicolumn{2}{|c|}{\rule{0pt}{2.5ex}\textbf{Application Layer}} \\ \hline
Video playback interval & 60 s \\ \hline
Number of independent runs & 30 \\ \hline
MPEG-DASH target buffer level & 30 s \\ \hline
Throughput estimation window & 50 ms \\ \hline
Playback buffer size & 512 MB \\ \hline

\multicolumn{2}{|c|}{\rule{0pt}{2.5ex}\textbf{Transport Layer}} \\ \hline
Initial congestion window & 10 packets \\ \hline
Slow-start threshold & 65535 bytes \\ \hline
Idle timeout & 30 s \\ \hline
Reordering threshold & 2 packets \\ \hline
Maximum number of tail loss probes & 5 \\ \hline
Minimum retransmission timeout & 200 ms \\ \hline
Delayed acknowledgment timeout & 25 ms \\ \hline
Initial RTT estimate & 333 ms \\ \hline

\multicolumn{2}{|c|}{\rule{0pt}{2.5ex}\textbf{Congestion Control Algorithms}} \\ \hline
TCP & BBR, NewReno, CUBIC \\ \hline
QUIC & BBR, NewReno \\ \hline

\multicolumn{2}{|c|}{\rule{0pt}{2.5ex}\textbf{MAC and Link Layer}} \\ \hline
Modulation and coding scheme & Adaptive based on CQI \\ \hline
Scheduler type & Padded hybrid beamforming \\ \hline
Number of resource blocks & 1 \\ \hline
Chunks per resource block & 72 \\ \hline
Number of RBs per RBG & 1 \\ \hline
Symbols per slot & 30 \\ \hline
Symbol period & 4.16 us \\ \hline
Subframe period & 100 us \\ \hline

\multicolumn{2}{|c|}{\rule{0pt}{2.5ex}\textbf{Physical Layer}} \\ \hline
Carrier frequency & 28 GHz \\ \hline
System bandwidth & 100.008 MHz \\ \hline
Subcarrier spacing & 28.9375 kHz \\ \hline
NTN channel model & 3GPP TR 38.811  \\ \hline
Terrestrial channel model & 3GPP TR 38.901 \\ \hline
Channel scenario & 3GPP Rural\\ \hline
UE antenna gain & 3 dBi \\ \hline
IAB node antenna gain & 40 dBi \\ \hline
Satellite antenna gain & 40 dBi \\ \hline
UE transmit power & 23 dBm \\ \hline
IAB transmit power & 40 dBm \\ \hline
Satellite transmit power & 40 dBm \\ \hline
Noise figure & 5 dB \\ \hline
Antenna model & Uniform planar array \\ \hline
\end{tabular}
\end{table}

 Table \ref{tab:simulation_parameters} shows different congestion control algorithms that are tested for our system model and used for comparison, and also summarizes the complete set of simulation parameters. The simulations are conducted with the primary objective of evaluating whether the considered configurations can reliably playback 60~s of video content. The User Equipment (UEs) are assumed to remain stationary during each simulation run, and the considered topology includes only a single terrestrial IAB node. The NTN channel model, 3GPP TR 38.811, does account for satellite-motion-induced channel effects, including Doppler shift and time-varying propagation characteristics. However, the serving satellite is assumed to remain in coverage for the full simulation duration. 
 

\begin{table}[t]
\centering
\caption{Interruption, latency, and fairness metrics}
\label{tab:performance_summary}

\renewcommand{\arraystretch}{1.18}
\setlength{\tabcolsep}{5pt}

\begin{tabular}{|p{0.23\columnwidth}|p{0.17\columnwidth}|p{0.17\columnwidth}|p{0.17\columnwidth}|}
\hline
\multicolumn{4}{|c|}{\rule{0pt}{2.5ex}\textbf{Interruption and Latency Metrics}} \\ \hline
\textbf{Algorithm} & \textbf{Interruption Duration (s)} & \textbf{Number of Interruptions} & \textbf{Latency (ms)} \\ \hline
QUIC-BBR & 2.94 & 42 & 12.65 \\ \hline
QUIC-NewReno & 34.33 & 55 & 7.48 \\ \hline
TCP-BBR & 6.39 & 328 & 28.86 \\ \hline
TCP-CUBIC & 0.00 & 50 & 289.42 \\ \hline
TCP-NewReno & 0.00 & 78 & 287.65 \\ \hline

\multicolumn{4}{|c|}{\rule{0pt}{2.5ex}\textbf{Fairness Metrics}} \\ \hline
\textbf{Algorithm} & \textbf{Throughput} & \textbf{Bitrate} & \textbf{Playback Duration} \\ \hline
QUIC-BBR & 0.270 & 0.349 & 0.909 \\ \hline
QUIC-NewReno & 0.249 & 0.455 & 0.682 \\ \hline
TCP-BBR & 0.179 & 0.228 & 0.930 \\ \hline
TCP-CUBIC & 0.309 & 0.355 & 0.995 \\ \hline
TCP-NewReno & 0.220 & 0.288 & 0.990 \\ \hline
\end{tabular}
\end{table}
\begin{figure*}[!t]
    \centering
    \includegraphics[width=0.88\textwidth]{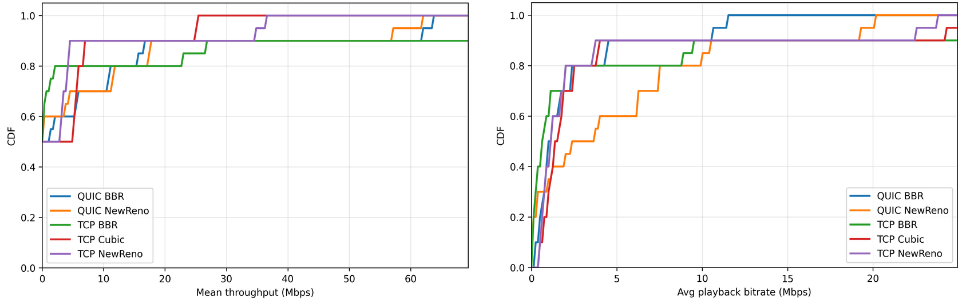}
    \caption{CDFs of average playback bitrate and mean throughput.}
    \label{fig:bitrate_throughput_cdf}
\end{figure*}

\subsection*{Simulation Results}

\subsubsection{Playback Continuity}
We first evaluate playback continuity, since the most fundamental requirement of a video streaming system is that users are able to complete the session with minimal stalling. For this purpose, we analyze playback duration, interruption duration, and number of interruptions. Playback duration indicates whether a user is able to watch the full 60~s video session. In the ideal case, playback duration reaches 60~s, which means that the client successfully sustains delivery for the complete playback interval. Any value below 60~s indicates that the session was interrupted and the playback could not be completed within the 60~s of simulation.

\begin{figure}[!t]
    \centering
    \includegraphics[width=1.0\columnwidth]{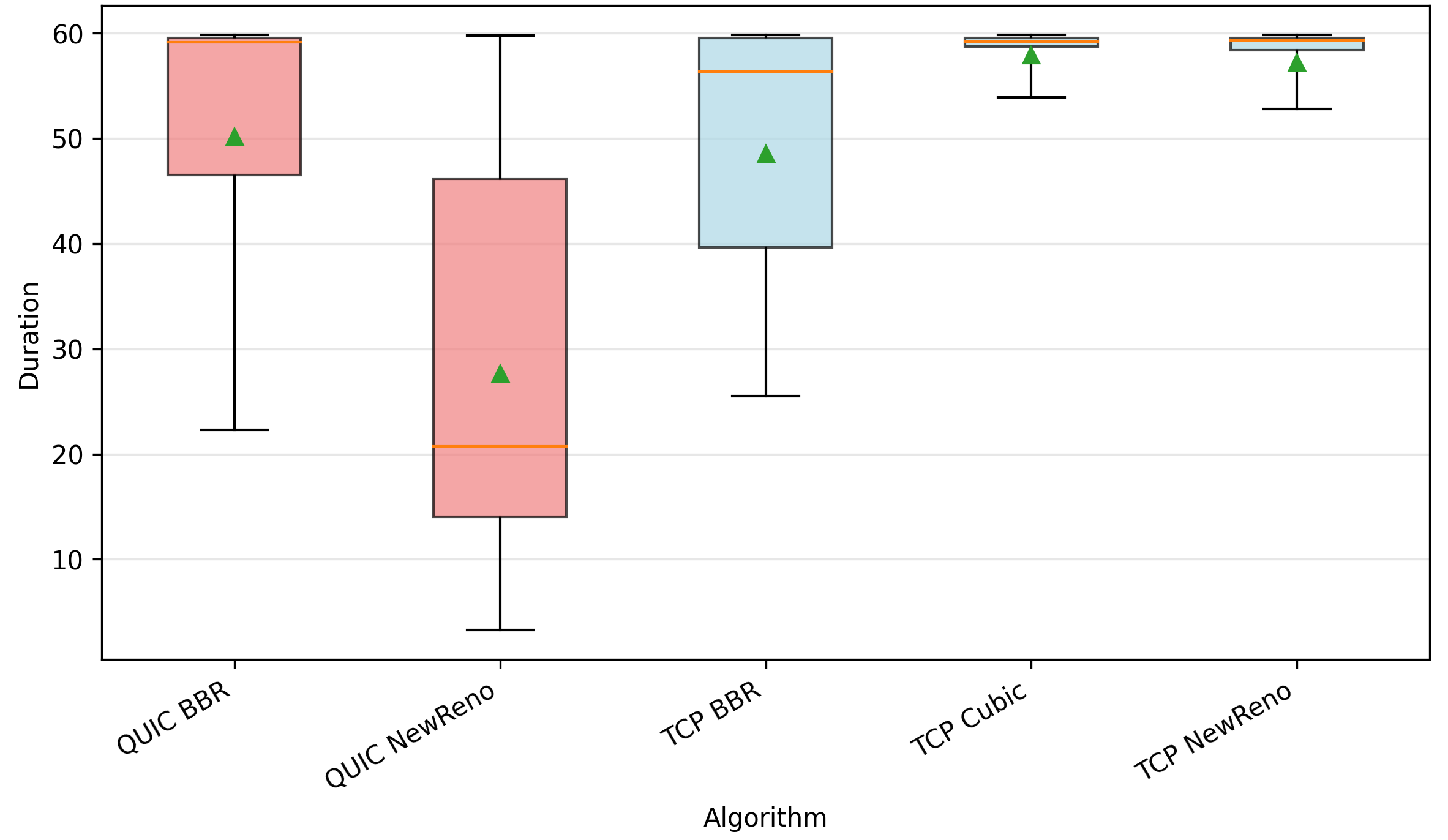}
    \caption{Playback duration.}
    \label{fig:playback_duration_boxplot}
\end{figure}

Fig.~\ref{fig:playback_duration_boxplot} presents the distribution of playback duration across all evaluated transport and congestion control configurations. Because outcomes differ from one user to another, a probability distribution is used to represent this variability across users. The simulation is run 30 times for each algorithm.
TCP-NewReno and TCP-CUBIC show the highest median playback durations, both close to the 60~s target. QUIC-BBR and TCP-BBR follow with slightly lower values, while QUIC-NewReno records the lowest median. Overall, the differences in playback duration are modest, and by themselves, do not suggest major differences in QoE.

Interruption duration captures the total time during which playback is stalled. For this metric, lower values are preferred, and the ideal case is 0s. The number of interruptions is measured using the total stall count. This metric reflects how often playback is disrupted. It is important because two schemes may have similar total interruption duration while producing very different viewing experiences if one causes frequent short stalls and the other causes only occasional interruptions. Lower values are therefore preferred, with zero interruptions representing uninterrupted playback.

Table~\ref{tab:performance_summary} reports the median interruption duration and the mean total number of stalls across repeated runs. TCP-CUBIC and TCP-NewReno each achieve a median interruption duration of 0~s, indicating stall-free playback for at least half of the users. QUIC-BBR shows a median interruption duration of 2.94~s, while TCP-BBR records a larger median of 6.39s. QUIC-NewReno exhibits the highest median interruption duration at 34.33~s. For the mean total number of stalls, QUIC-BBR yields the lowest value at 42, followed by TCP-CUBIC at 50 and QUIC-NewReno at 55. TCP-NewReno shows a higher mean stall count of 78, while TCP-BBR performs worst with a mean of 328 stalls. An important observation is that although QUIC-BBR and QUIC-NewReno have relatively similar stall counts, their interruption durations differ substantially, which suggests that the key distinction between these two schemes lies in stall severity rather than stall frequency. Taken together, the interruption metrics differentiate the evaluated schemes much more clearly than playback duration alone.

\subsubsection{Latency}
We next consider latency, which reflects the responsiveness of the end-to-end path. Latency is defined as one-half of the measured round-trip time (RTT), representing the approximate time taken for a packet to travel one way to the receiver.

Table~\ref{tab:performance_summary} reports the median latency values for each transport and congestion control configuration. The QUIC-based schemes yield the lowest median latency values, with QUIC-NewReno and QUIC-BBR achieving 7.48~ms and 12.65~ms, respectively. TCP-BBR follows at 28.86~ms. In contrast, TCP-CUBIC and TCP-NewReno exhibit substantially larger median latencies of 289.42~ms and 287.65~ms. These results indicate that the choice of transport and congestion control has a strong effect on end-to-end responsiveness even when some playback-related metrics remain relatively close.

\subsubsection{Playback Quality and Transport Efficiency}
After establishing continuity and responsiveness, we examine average playback bitrate and mean throughput to assess the delivered media quality and the transport efficiency supporting that quality. For both CDF plots, curves located further to the right correspond to higher values and therefore better performance.

Fig.~\ref{fig:bitrate_throughput_cdf} shows the CDFs of average playback bitrate and mean throughput. In the bitrate distribution, QUIC-NewReno attains the highest median playback bitrate at approximately 3.28~Mbps and also exhibits a longer upper tail than the other schemes, indicating that some runs achieve substantially higher delivered quality. TCP-CUBIC, TCP-NewReno, and QUIC-BBR occupy an intermediate range, with median playback bitrates of 1.41~Mbps, 1.16~Mbps, and 1.03~Mbps, respectively. TCP-BBR records the lowest median playback bitrate at 0.72~Mbps. Overall, the bitrate distributions are spread across a wide range, indicating noticeable variability in the delivered video quality across runs.

In the throughput distribution of Fig.~\ref{fig:bitrate_throughput_cdf}, TCP-CUBIC achieves the highest median throughput at 2.46~Mbps, followed by TCP-NewReno at 1.50~Mbps and QUIC-BBR at 1.24~Mbps. QUIC-NewReno has a lower median throughput of 0.50~Mbps, while TCP-BBR shows the lowest median throughput at 0.04~Mbps. The throughput distributions are highly dispersed, particularly for TCP-BBR, where the sharp rise near very small values and the long upper tail indicate that most runs experience low typical throughput while a small number of runs achieve much larger values.

\subsubsection{Fairness}
Finally, we analyze fairness using Jain's fairness index to determine how evenly the available resources are distributed across users, since strong aggregate performance may still hide unequal user experiences. For a set of $N$ users with values $x_1, x_2, \ldots, x_N$, the fairness index is defined as
\begin{equation}
J(\mathbf{x}) = \frac{\left(\sum_{i=1}^{N} x_i\right)^2}{N \sum_{i=1}^{N} x_i^2},
\label{eq:jain_fairness}
\end{equation}
where $x_i$ represents the value of the metric for user $i$, such as throughput, playback bitrate, or playback duration. The value of $J(\mathbf{x})$ approaches 1 when the allocation is highly fair and decreases as the distribution becomes more uneven.

Table~\ref{tab:performance_summary} shows that playback duration fairness remains high for most configurations, particularly for TCP-CUBIC and TCP-NewReno. However, the fairness values for throughput and playback bitrate are substantially lower across all schemes. This indicates that even when users achieve broadly similar playback durations, the underlying allocation of throughput and delivered video quality is much less balanced.

\subsection*{Discussion}
The results indicate that no single transport and congestion control combination is optimal across all metrics. Instead, each scheme reflects a different tradeoff among playback continuity, bitrate delivery, throughput, latency, and fairness. Among the evaluated configurations, QUIC-BBR provides the most balanced overall performance. It combines competitive playback duration with shorter interruption periods and substantially lower latency than the TCP loss-based variants. These characteristics make it well suited for video streaming scenarios where both continuity and responsiveness are important. TCP-CUBIC and TCP-NewReno achieve strong playback duration and throughput, but their much higher latency reduces their suitability for delay-sensitive applications. QUIC-NewReno achieves high playback bitrate and low latency, but its significantly larger interruption duration weakens its overall QoE performance.

\section*{Conclusion}
We have evaluated several transport and congestion control configurations for adaptive video streaming over a LEO-backhauled 5G IAB network. The results show that the tested schemes exhibit distinct performance tradeoffs, with no single approach dominating every metric. Among them, QUIC-BBR achieves the best overall balance across playback continuity, latency, and delivery performance, making it the most suitable option in the evaluated scenario. The study presented in this paper can be extended by exploring the following directions:

\begin{itemize}
    \item Extend the framework to incorporate UE mobility and constellation-level satellite dynamics, so that the impact of handover and possible outage events on transport and MPEG-DASH QoE can be evaluated.
    
    \item Develop adaptive or hybrid mechanisms that dynamically select or tune congestion control behavior based on real-time application requirements and network conditions.
    \item Incorporate cross-layer information, such as link variability and handover events, to enable more informed transport-layer adaptations.
    \item Integrate generative AI models to analyze complex network patterns and automate the selection of optimal protocol configurations.
    \item Leverage observed per-metric QoE tradeoffs to design targeted transport and streaming optimizations for diverse traffic profiles.
\end{itemize}

\ifCLASSOPTIONcaptionsoff
\newpage
\fi

\bibliographystyle{IEEEtran}
\bibliography{references}

\begin{IEEEbiographynophoto}{Muhammad Adeel Zahid} is currently pursuing the M.Sc. degree with the Department of Electrical and Computer Engineering at the University of Manitoba, Canada. His research interests include cellular networks, transport-layer design, congestion control, and end-to-end network optimization.
\end{IEEEbiographynophoto}

\begin{IEEEbiographynophoto}{Ekram Hossain}(Fellow, IEEE) is a Professor and the Associate Head (Graduate Studies) of the Department of Electrical and Computer Engineering, University of Manitoba, Canada. He is a Member of the College of the Royal Society of Canada, and a Fellow of the Canadian Academy of Engineering and the Engineering Institute of Canada. His research interests include modeling, analysis, and optimization of 6G networks and beyond.
\end{IEEEbiographynophoto}

\begin{IEEEbiographynophoto}
{Peng Hu} (Senior Member, IEEE) is an Associate Professor in the Department of Electrical and Computer Engineering at the University of Manitoba, Canada. His research interests include non-terrestrial networks, space-air-ground integrated network systems, and network resilience.
\end{IEEEbiographynophoto}
\end{document}